# GIGAPYX SENSOR PERFORMANCE IN SPACE ENVIRONMENTS

Julien Michelot*[a], Maurin Douix[a], Jean-Baptiste Mancini[a], Marie Guillon[a], Kevin Melendez[b], Clément Ravinet[b], Mikael Jouans[b], Guy Estaves[b], Ronan Marec[b], Stéphane Demiguel[b], Alex Materne[c], Cédric Virmontois[c]

*Corresponding author : julien.michelot@pyxalis.com

[a]Pyxalis, 170 Rue de Chatagnon, 38430 Moirans, France; [b]Thales Alenia Space, 100 boulevard du midi, 06156 Cannes la Bocca, France; [c] Centre national d'études spatiale, 18 avenue Edouard Belin, 31401 Toulouse, France

## ABSTRACT

We present the results of the GIGAPYX-4600 image sensor in space environment, more specifically under different types of irradiations (protons and heavy ions). The GIGAPYX-4600 is a state-of-the-art 46M pixel multi-purpose, back-side illuminated CMOS image sensor. The sensor features high-speed (200 fps), low-noise (< 2e- rms), rolling shutter readout. It has been fabricated using 65 nm node CMOS technology, making use of capacitive deep trench isolation, thus exhibiting good MTF as well as excellent dark current characteristics. The GIGAPYX image sensor family is meant to be easily scalable thanks to a novel use of stitching technology. The assessed sensor features an impressive 46 Mpixels, but the sensor family is meant to be scaled up to 220 M pixels. The idea of this study was to investigate the radiation hardness of a commercially available off-the-shelf (COTS) image sensor that could be suitable for space-borne applications, such as earth observation or satellite vicinity surveillance. In the near future a radiation hard readout electronic for this sensor family will be made available off-the-shelf by Pyxalis.

This presentation will overview the different sensor key performances evolutions after proton irradiation up to a total fluence of 2.3e11 p+/cm² (62 MeV): dark current, dark current non-uniformity (DCNU), noise, non-linearity, saturation charge and photo-response non-uniformity (PRNU). As expected, degradations mostly occur on the dark current, DCNU and temporal noise. The orders of magnitude of the degradation are in the range of the already published high performance CIS technologies. Results obtained from heavy ions irradiations will demonstrate that the GIGAPYX is not only latch-up free at least up to 57 Mev.cm2/mg, but resistant to the blooming effects induced by a SET at pixel level thanks to its capacitive deep trench isolations. Various SEE effects have been studied, demonstrating encouraging results for such COTS device in order to fly, in particular in GEO orbit. All these radiations results will be used as inputs in designing space camera based on the GIGAPYX sensors, compatible with multiple-mission types.

## 1. INTRODUCTION

Pyxalis, a French-based image sensor design house, recently introduced the GIGAPYX image sensor family. These image sensors are notable for their very high resolution, low noise, and high frame rate capabilities [1]. They can be used in space borne applications such as Earth observation and space situational awareness. Although this family of sensors is not specifically designed for space environments, this paper focuses on the performance of the GIGAPYX-4600 sensor under heavy-ion radiation and total non-ionizing dose results.

Section 2 will focus on the image sensor architecture and technology, explaining how it has been designed so that the entire family can be manufactured using a single mask set. We will provide some key performance metrics of the image sensor prior to any irradiation. In Section 3, we will focus on the effect of total non-ionizing dose result and its impact on the sensor's performance. Section 4 will be dedicated to heavy-ion irradiation and its results. Finally, section 5 will conclude with respect to the usage of such devices for space-borne applications.

## 2. GIGAPYX SENSOR ARCHITECTURE AND TECHNOLOGY

Technology

GIGAPYX sensors are backside illuminated (BSI), rolling shutter, CMOS image sensors manufactured using an imager-dedicated CMOS technology offered by a French foundry on 300 mm diameter wafers. This technology is based on a 90 nm CMOS node and provides state-of-the-art features for high-end image quality. Among these features, it is worth mentioning:

- A pinned photodiode and its associated transfer gate, allowing lag-less true correlated double sampling (CDS) and thus permitting excellent readout noise in darkness. This type of photodiode natively has a surface implant used as a pinning layer that also acts as silicon surface state passivation, resulting in excellent measured dark current.

- Capacitive deep trench isolation (CDTI), which provides a physical barrier between neighboring pixels [2]. Thus, photo-generated electrons cannot travel from one pixel to another, significantly improving crosstalk and modulation transfer function (MTF).

- Four conductive copper layers that allow the manufacturing of large sensors, as the low resistivity of copper facilitates the transport of electrical signals across the chip.

- The possibility to use 2D-stitching techniques to produce sensors that are (much) larger than the mask reticle size.

The GIGAPYX sensor family philosophy relies on the principle of manufacturing different sensor resolutions using a unique mask set. Figure 1 shows how this is achieved: The mask reticle embeds the 9 necessary blocks to produce a GIGAPYX sensor. By photo-replicating blocks A, B, and C by a desired amount, different resolutions of image sensors are achieved.

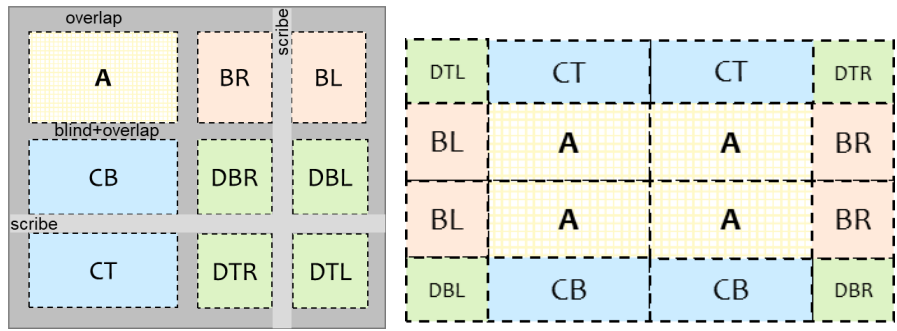

Figure 1. Schematic overview of the mask reticle (left), Sensor layout after photo-replication (right).

Table 1 summarizes the available formats using a single mask reticle. The first sensor out of this family being produced is GIGAPYX-4600, which is also the sensor under test in the following paragraphs.

Table 1 : Size features of the different image sensors available in the GIGAPYX family

| GIGAPYX Family | Format | block A along X and Y axes | Definition (Width x Height) | Matrix sizes in mm (Width x Height) | Diagonal in mm |
|---|---|---|---|---|---|
| GIGAPYX-1400 |  | 2, 3 | 4160 x 3256 | 18.3 x 14.5 | 23.35 |
| GIGAPYX-2700 | Super 35 mm | 3, 4 | 6240 x 4356 | 27.5 x 19.4 | 33.65 |
| GIGAPYX-3700 |  | 4, 4 | 8320 x 4356 | 36.6 x 19.4 | 41.42 |
| GIGAPYX-4600 | 35 mm Full frame | 4, 5 | 8320 x 5456 | 36.6 x 24.2 | 43.87 |
| GIGAPYX-8000 |  | 5, 7 | 10400 x 7656 | 45.8 x 33.9 | 56.98 |
| GIGAPYX-8200 | 65 mm | 6, 6 | 12480 x 6556 | 54.9 x 29 | 62.08 |
| GIGAPYX-11000 |  | 6, 8 | 12480 x 8756 | 54.9 x 38.7 | 67.17 |
| GIGAPYX-15100 | 65 mm square | 6, 11 | 12480 x 12056 | 54.9 x 53.2 | 76.45 |
| GIGAPYX-22000 | Max size | 8, 12 | 16640 x 13156 | 73.2 x 58.1 | 93.45 |

Sensor architecture

GIGAPYX sensors embed 4.4 µm pitch high dynamic range (HDR) pixels. To extend pixel dynamic range, these pixels utilize a dual gain architecture, thanks to an additional capacitor that can be connected to the sensing node through a dedicated transistor (see Figure 2) [3]. The sensor can operate in three different modes:

- Single gain readout, either in high gain (HG) or low gain (LG), up to 150 frames per second for a full-frame readout.

- HDR readout: The two gains are used simultaneously without any reset of the sensing node, thus after HDR reconstruction, intra-scene dynamic range is significantly improved, reaching about 90 dB. Since two analog-to-digital conversions must be carried out and two images are produced, the output frame rate is reduced to 70 fps in this readout mode.

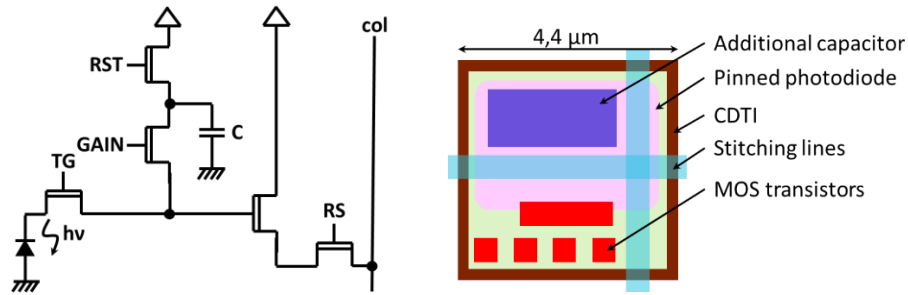

Figure 2. Schematic overview of the pixel architecture (left), pixel layout top view (right). As can be seen, the two stitching lines occupy a non-negligible part of the pixel layout. Since the pixel is backside-illuminated, the additional capacitor can be placed directly over the photodiode without hindering the impinging light.

The sensor architecture block diagram can be found in Figure 3. Pixel data is converted to digital values by column-parallel ADCs placed on the top and bottom sides of the sensor. Consequently, data output can also be found on the top and bottom sides of the sensor. Line decoders are located in the blocks on the left and right of the pixel array. Other general-purpose blocks, such as temperature sensors, biasing circuits, and timing generators, can be found in the corners.

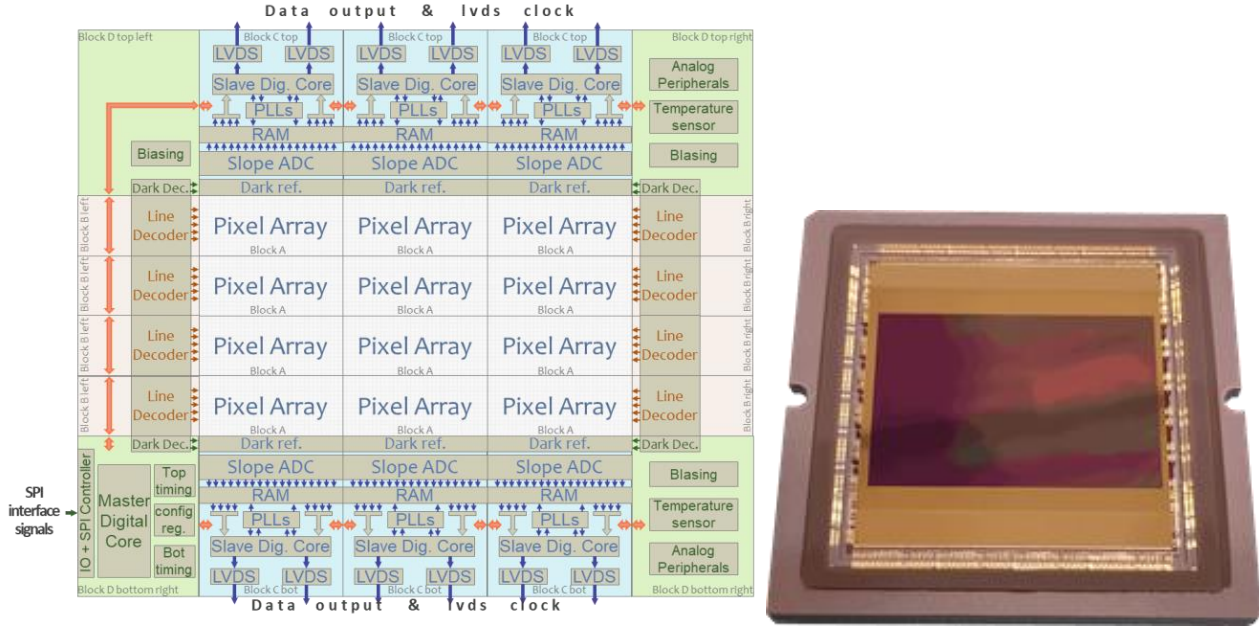

Figure 3. GIGAPYX sensor bloc diagram (left), GIGAPYX-4600 sensor (right).

Pre-irradiation electro-optical performances

Typical sensor electro-optical performances can be found in Table 2.

Table 2: Measured sensor electro-optical performances

| Parameter | Typical Value | Unit |
|---|---|---|
| Linear full-well (HG / LG) | 5 / 50 | ke- |
| Conversion factor (HG / LG) | 188 / 18 | µV/e- |
| Temporal Noise in darkness (HG) | 1.6 | e- RMS |
| Linearity Error, 5-95% QSATLIN | 2 | % |
| Dynamic Range, HG | 70 | dB |
| Dynamic Range, HDR | 90 | dB |
| Peak QE x FF, at 560 nm (mono) | > 72 | % |
| Dark Current at 35°C | 7 | e-/s |
| | 25 | pA/cm² |
| Fixed-Pattern Noise (FPN) | < 0.4 | % |
| Photo-Response Non-Uniformity (PRNU) | < 2 | % |
| Power-consumption, full operation at 150 fps | < 6 | W |

# 3. TOTAL NON-IONIZING DOSE TESTS

TNID tests were performed using three sensors. A reference sensor was not exposed to any proton flux, whereas the two other sensors were exposed to two different irradiation doses. These sensors were unbiased (grounded) during the tests, the glass lid was removed from the sensors, and tests were carried out at room temperature. The tests occurred in May 2023 at Université Catholique de Louvain (UCL) in Belgium. Table 3 provides the irradiation conditions for the two irradiated samples.

Table 3: Irradiation condition for the two irradiated samples

| Sample ID | Proton Energy (MeV) | Total non-ionizing dose (TNID) in rads | Fluence in p/cm² |
|---|---|---|---|
| 17 | 62 | 10200 | 7.6e10 |
| 24 | 62 | 30880 | 2.3e11 |

We observed no specific degradation in pixel gain or full well capacity; the maximum observed variation was 3%, which we attribute to measurement uncertainty. The observation was similar for non-linearity and PRNU performance.

On the other hand, readout noise in darkness degraded with TNID:

- For circuit 17 (10 krad) readout noise increased from 1.6 e- rms (HG) and 17.4 e- rms (LG) to 2.3 e- rms (HG) and 20.4 e- rms (LG) after irradiation.

- For circuit 24 (30 krad) redout noise increased from 1.7 e- rms (HG) and 19.3 e- rms (LG) to 3.3 e- rms (HG) and 26.8 e- rms (LG) after irradiation.

Dark current and DCNU also increased after irradiation, figure 4 gives the dark current histograms of the chip before irradiation, after 10 krad irradiation and after 30 krad irradiation. This dark current was measured at an operating temperature of 31 °C. Dark current mean value as well as dark current doubling temperature are provided in Table 4.

Table 4: Dark current figures for reference samples and irradiated samples

| Sample ID | Median dark current at 31 °C (pA/cm²) | Mean DCNU at 31 °C (pA/cm² rms) | Dark current doubling temperature in °C |
|---|---|---|---|
| Reference | 3.1 | 25 | 7.2 |
| 17 | 52.5 | 2065 | 7.5 |
| 24 | 208.5 | 3192 | 7.5 |

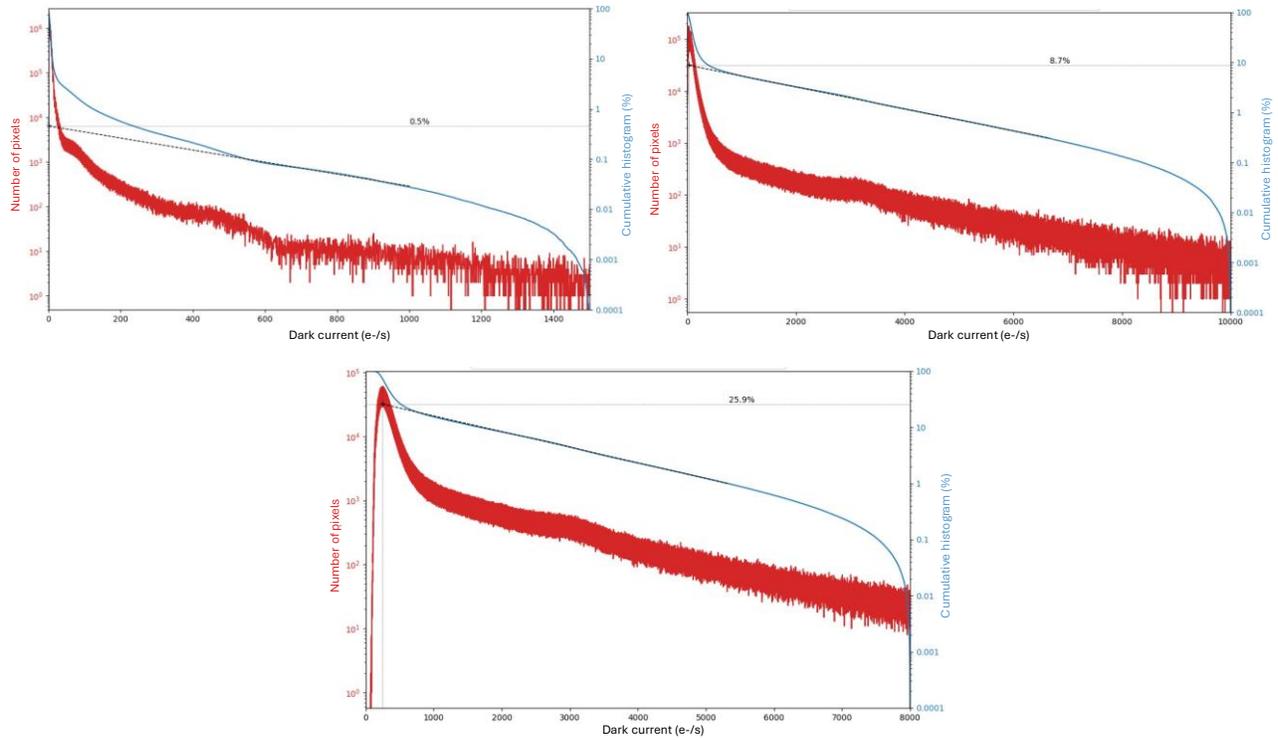

Figure 4. Dark current histograms before irradiation (top left), after 10 krad irradiation (sample 17, top right) and after 30 krad irradiation (sample 24, bottom).

## 4. HEAVY IONS TESTS

Heavy ion tests have been completed at RADEF facility located in the Accelerator Laboratory at the University of Jyväskylä, Finland. The experimental setup is described underneath. It consisted in assessing Single Event Latchp (SEL) sensitivity and various Single Event Effect (SEE) through dynamic test. The irradiations tests are performed in a closed irradiation room (in air mode), which includes remote-controlled component movement apparatus and collimator, and ion beam diagnostic equipment for real-time analysis of beam quality and intensity. The control station for the RADEF equipment and the user work station are located outside of the experiment cave. A series of interfaces are available to link the devices under test to the user's test equipment.

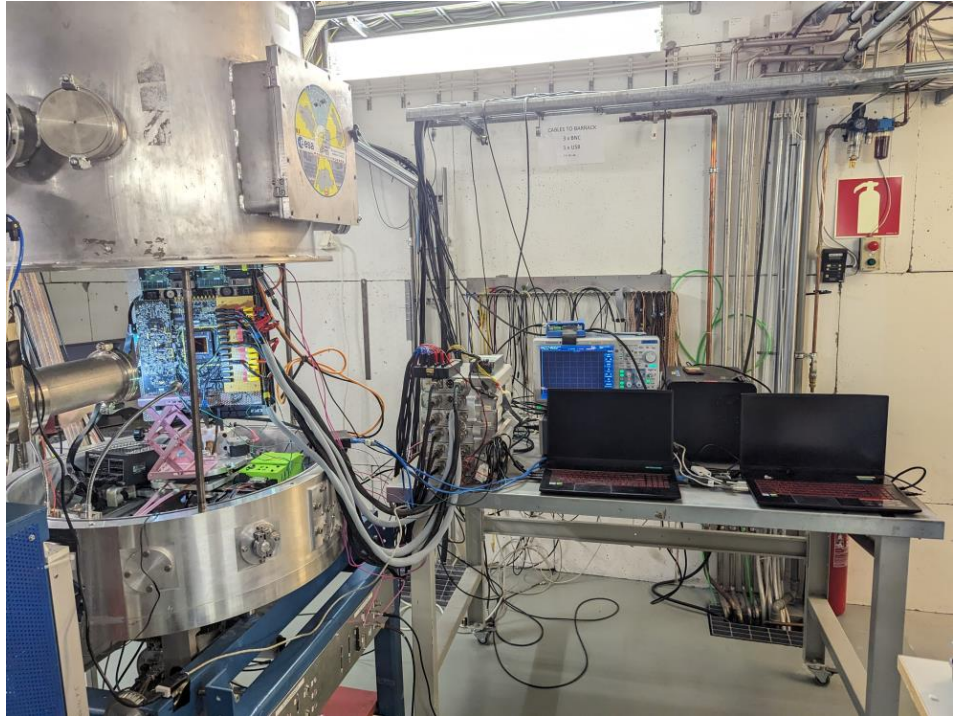
Figure 5. Heavy Ions experiment test setup prepared at RADEF.

The radiation test board has been developed by Pyxalis. Additional test means, provided by Thales Alenia Space have been connected to it including:
- A specific test software to monitor Sensor registers connected to each FPGA through dedicated serial lines. A processing software collects time stamped registers values sent by each FPGA, and computes statistics on register events,
- A delatcher system (TALIS) to power up the CMOS sensor, and monitor the current on specific dedicated lines to detect any latch up during the tests,
- A temperature control system added to set the CMOS sensor at the right temperature during latch up tests.

Pyxalis also provided additional test means to capture the images sent by the FPGA, save them on a hard drive, analyze and post-process the images to detect and characterize the defects. The test setup is depicted below.

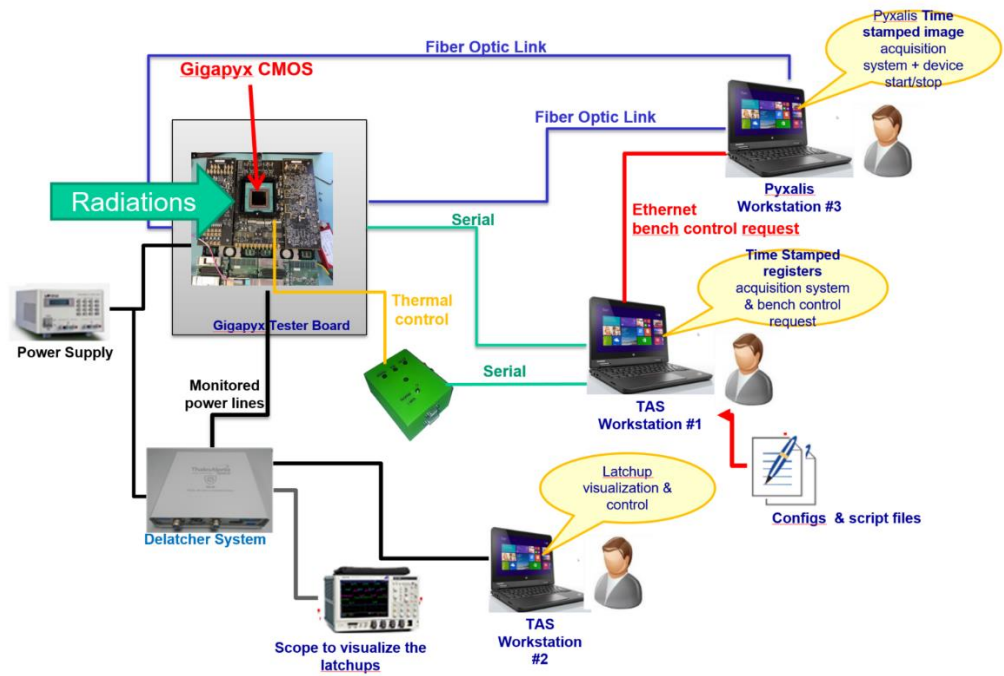

Figure 6. GIGAPYX SEE test setup schematic.

The system is composed of 2 FPGA boards, 1 DUT board and 1 Sensor Board onto which the CMOS Sensor to test was mounted.

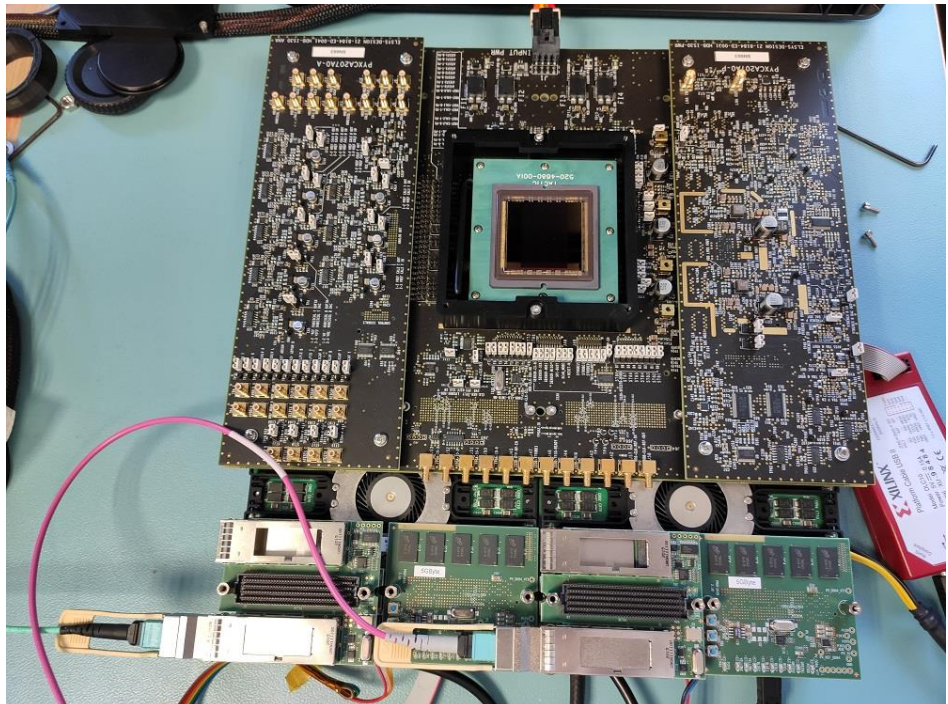

Figure 7. GIGAPYX SEE test board.

The temperature targeted was near 60°C for the latch up tests and kept as low as possible near the normal operating temperature for the dynamic tests.

A specific test hardware and software have been designed for latch-up detection, protection and logging. This specific equipment, called TALIS (Thales Anti latch-up Integrated System) allows to perform Single Event Latch up (SEL) characterization, while protecting the device under test against damage or destruction due to thermal effects. These tests were performed at maximum operating voltage. All relevant power lines for the SEL test were powered by the TALIS, which continuously monitored the current consumption. If a threshold current defined initially was exceeded, it was considered as an event and potentially a latch up. The data recorded by TALIS confirmed or not the occurrence of a SEL event.

In the present test campaign, no latch up have been detected up to the performed test range of 57 Mev.cm$^2$/mg. But if there were a SEL event detection, the power supply would have been maintained during a defined 'Hold Time', switched off during a defined 'Off Time' and then restarted with the nominal expected current consumption. At the same time, a trigger would have been sent to the oscilloscope to capture the corresponding current waveform in order to verify if was a SEL or rather a SEFI signature.

The dynamic test to characterize the SEE operates as follow:
- On each FPGA board, a serial line is available. The FPGA firmware running on each FPGA sends periodically the time stamped register values of the CMOS Sensor. A TAS test software (SW) to these serial lines gathers the registers values and makes on live processing of the register events.
- From each FPGA, the time stamped images are sent periodically to the Pyxalis image acquisition system. This time stamp has the same reference time than the registers time stamp in order to perform offline correlation based on time between register events and image events.

The test SW developed by TAS can drive the test setup, to set the BEAM ON/OFF, to power ON/OFF the boards, to periodically acquire the register values provided by the FPGA and reset registers configuration. It will check the events inside the registers, also on image mean value computed. In case an error event is detected on a critical register or mean value outside of programmable tolerance, the SW will shut down the BEAM, power OFF/ON (or RESET) the boards, restart the test and set the BEAM ON again.

The goal of these dynamic tests was to determine the signature of the events occurring during irradiation, and their cross-section (i.e. ratio of the number of events to the total particle fluence in ions/cm²) versus LET. The fluence used during the test were between $10^6$ ions/cm² and $10^7$ ions/cm². However, the run may be stopped before $10^6$ ions/cm² only for LET values < 60 MeV.cm²/mg, when enough number of events (i.e. more than 100) have been recorded. For each type of event, the cross section was determined on 2 devices.

The SEU register test was done in 4 different configurations mode (high gain and HDR, in pixel mode and with pattern generator). There is no difference on the saturated cross section between all configurations. Single/multiple bit event upset and SEFI have been characterized on the register, as well SEFI on the CMOS panel (i.e. SEFI seen in the ADC, corrupting the image and requiring a sensor reset). Some pixels events occurred but they were resistant to any blooming effects induced by a SET at pixel level thanks to its deep trench isolations. The plotted cross section curves and Weibull fit parameters were very clean. The SEFI one related are represented below.

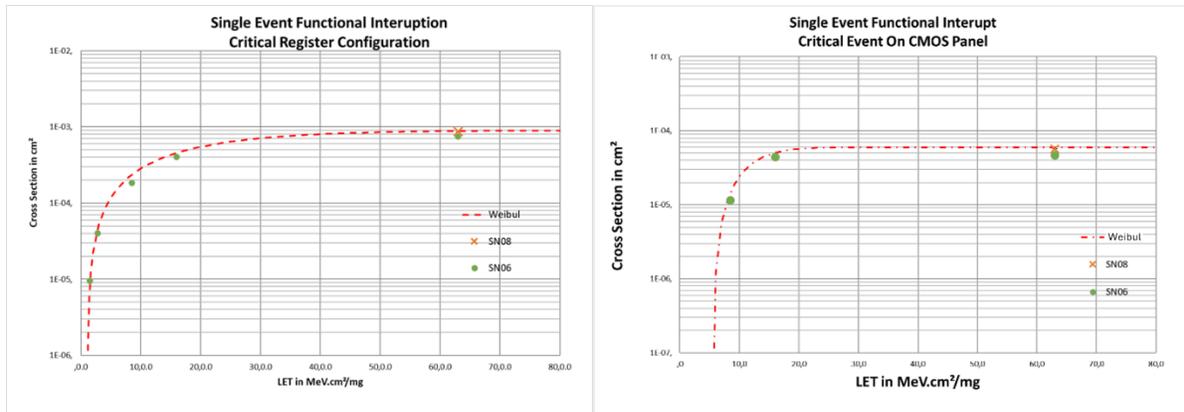

Figure 8. SEFI on critical register and CMOS panel cross section and Weibull fit parameters

Thanks to these measurements, associated events occurrence has been calculated for two orbit (GEO/LEO) and reported in the table below.

Table 5: Evaluation of the occurrence rate for different types of single events and for 2 types of orbits.

| Events due to heavy ions | GEO Orbit Event/device/day | LEO Orbit Event/device/day |
|---|---|---|
| SEU Registers | $3{,}32\times10^{-3}$ | $1{,}17\times10^{-3}$ |
| MBU Registers | $9{,}36\times10^{-4}$ | $1{,}65\times10^{-4}$ |
| SEFI on Critical registers configuration | $5{,}66\times10^{-3}$ | $1{,}29\times10^{-4}$ |
| SEFI on Critical events on CMOS Panel | $2{,}36\times10^{-4}$ | $5{,}29\times10^{-5}$ |

The sensitivity on ADC has been demonstrated relatively low. However, the registers exhibit more numerous events and lower LET threshold which would require further SEE characterizations to assess the sensitivity to protons to refine the performance in LEO orbit. In any case, the results are encouraging for such COTS device in order to fly, in particular in GEO orbit.

## 5. CONCLUSION

We have presented the results of the multi-purpose GIGAPYX-4600 image sensor in protons and heavy ions environment to assess the radiation hardness of such COTS image sensor. The conclusion is that it could be suitable for space-borne applications, such as earth observation or satellite vicinity surveillance.

The different sensor key performances evolutions after proton irradiation up to a total fluence equivalent to 30 krad have been overviewed. As expected, degradations mostly occurred on the dark current, DCNU and temporal noise, while the other performance (non-linearity, saturation charge and PRNU) remain unchanged. The orders of magnitude of the degradation are in the range of the already published high performance CIS technologies [4], [5].

Results obtained from heavy ions irradiations demonstrated that the GIGAPYX-4600 sensor is not only latch-up free at least up to 57 Mev.cm²/mg, but resistant to the blooming effects induced by a SET at pixel level thanks to its deep trench isolations. Various SEE effects have been studied, demonstrating encouraging results for such COTS device in order to fly, in particular in GEO orbit. All these radiations results will be used as inputs in designing space camera based on the GIGAPYX family sensors, compatible with multiple-mission types. In a near future radiation hard readout electronics for this sensor family will be made available off-the-shelf by Pyxalis.

## ACKNOWLEDGEMENTS

These results have been achieved in a frame of R&T, co-founded by CNES and Thales Alenia Space.

## REFERENCES

[1] M. Dubois et al., "GIGAPYX, a very high-resolution CIS platform," ESA-ESTEC Space & Scientific CMOS Image Sensors Workshop 2022
[2] N. Ahmed et al., "MOS Capacitor Deep Trench Isolation for CMOS image sensors," 2014 IEEE International Electron Devices Meeting, San Francisco, CA, USA, 2014
[3] K. D. Stefanov and M. J. Prest, "Pinned Photodiode Imaging Pixel With Floating Gate Readout and Dual Gain," in IEEE Transactions on Electron Devices, vol. 70, no. 6, pp. 3136-3139, June 2023
[4] C. Virmontois et al., "Displacement Damage Effects Due to Neutron and Proton Irradiations on CMOS Image Sensors Manufactured in Deep Submicron Technology," in IEEE Transactions on Nuclear Science, vol. 57, no. 6, pp. 3101-3108, Dec. 2010
[5] V. Goiffon et al., "Radiation Effects in Pinned Photodiode CMOS Image Sensors: Pixel Performance Degradation Due to Total Ionizing Dose," in IEEE Transactions on Nuclear Science, vol. 59, no. 6, pp. 2878-2887, Dec. 2012